\documentclass[aip,apl,reprint,superscriptaddress]{revtex4-1}

\pdfoutput=1

\usepackage{amsmath}%
\usepackage{amsfonts}%
\usepackage{amssymb}%
\usepackage{graphicx}

\begin{document}

\title{Ion Traps Fabricated in a CMOS Foundry}

\author{K.~K.~Mehta}
\thanks{K.~K.~Mehta, A.~M.~Eltony, and C.~D.~Bruzewicz contributed equally to this work.}
\affiliation{Department of Electrical Engineering and Computer Science, Massachusetts Institute of Technology, Cambridge, Massachusetts 02139, USA}

\author{A.~M.~Eltony}
\thanks{K.~K.~Mehta, A.~M.~Eltony, and C.~D.~Bruzewicz contributed equally to this work.}
\affiliation{Center for Ultracold Atoms, Research Laboratory of Electronics and Department of Physics, Massachusetts Institute of Technology, Cambridge, Massachusetts 02139, USA}

\author{C.~D.~Bruzewicz}
\thanks{K.~K.~Mehta, A.~M.~Eltony, and C.~D.~Bruzewicz contributed equally to this work.}
\affiliation{Lincoln Laboratory, Massachusetts Institute of Technology, Lexington, Massachusetts 02420, USA}

\author{I.~L.~Chuang}
\affiliation{Center for Ultracold Atoms, Research Laboratory of Electronics and Department of Physics, Massachusetts Institute of Technology, Cambridge, Massachusetts 02139, USA} 

\author{R.~J.~Ram}
\affiliation{Department of Electrical Engineering and Computer Science, Massachusetts Institute of Technology, Cambridge, Massachusetts 02139, USA}

\author{J.~M.~Sage}
\email[]{jsage@ll.mit.edu}
\affiliation{Lincoln Laboratory, Massachusetts Institute of Technology, Lexington, Massachusetts 02420, USA}

\author{J.~Chiaverini}
\email[]{john.chiaverini@ll.mit.edu}
\affiliation{Lincoln Laboratory, Massachusetts Institute of Technology, Lexington, Massachusetts 02420, USA}

\date{\today}

\begin{abstract}
We demonstrate trapping in a surface-electrode ion trap fabricated in a 90-nm CMOS (complementary metal-oxide-semiconductor) foundry process utilizing the top metal layer of the process for the trap electrodes.  The process includes doped active regions and metal interconnect layers, allowing for co-fabrication of standard CMOS circuitry as well as devices for optical control and measurement.  With one of the interconnect layers defining a ground plane between the trap electrode layer and the p-type doped silicon substrate, ion loading is robust and trapping is stable.  We measure a motional heating rate comparable to those seen in surface-electrode traps of similar size.  This is the first demonstration of scalable quantum computing hardware, in any modality, utilizing a commercial CMOS process, and it opens the door to integration and co-fabrication of electronics and photonics for large-scale quantum processing in trapped-ion arrays.
\end{abstract}

%\pacs{}% insert suggested PACS numbers in braces on next line

\maketitle %\maketitle must follow title, authors, abstract and \pacs

Trapped atomic ions are a promising system for large-scale quantum processing\cite{RevModPhys.75.281,Blatt:Wine:Nat08}, as all required basic quantum operations have been demonstrated with low error\cite{Inns:HiFi2qubit:NatPhys:08,Oxford:HiFiRdout:PRL:08,NIST:HifiMicrogate:12}.  However, these demonstration experiments typically consist of relatively few ions ($\lesssim10$) manipulated with optical beams and electronic signals routed from outside the ion-trap vacuum chamber.  In order to scale the system to the number of quantum bits (qubits) required to provide speedups over classical computing methods, trap arrays holding orders of magnitude more ions are necessary.  Additionally, each array site will require local control, readout electronics, and optics for scalability.

Current microfabricated ion traps (and, in fact, realizations of any scalable quantum processing technology) depend on specialized, nonstandard processes in research clean-room facilities.  The traps are often built upon non-silicon substrates\cite{NIST:SET:PRL:06,Chaung:CryoHeatRates:PRL08}, and where silicon is used, only a few metal layers (four maximum) have been implemented\cite{britton:nist:2008,SMIT:QIC:2009,Sandia:Bksideloading:arXiv:10,Oxford:CaCharging:APB:11,npl:monolithictrap:2012,NJP:GTRI:xjunctrap:2013,Innsbruck:CryoSiTrap:arXiv:14,sussex:2dhexrap:2014}.  None of the traps made on silicon substrates to date have had doped, active device fabrication available, and due to the idiosyncratic process steps used, the lithographic resolution is typically limited.  Repeatability at different facilities is almost impossible due to local process variations and substrate processing capabilities.

Here we describe the design and operation of an ion trap built into a standard high-resolution CMOS fabrication process.  Based on industry-standard practices and materials, there are active and passive layers beneath the trap-electrode layer that may enable integration of electronics and photonics\cite{shepardAPD:APL:2010,elecphotintegration:Orcutt:2012} for control and readout of trapped-ion quantum states.  Standardization of the foundry process permits any group to produce identical devices with high yield.

\begin{figure}[tbp]
\includegraphics[width=\columnwidth]{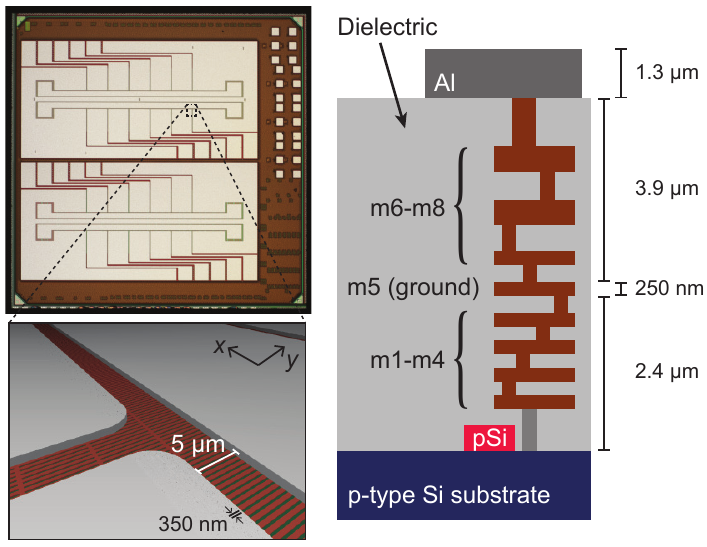}
\caption{(Color online) Die and process cross-section.  A micrograph of the fabricated $3\times3$~mm$^{2}$ die is shown in the upper left panel. The lower left panel shows a perspective rendering of the top aluminum trap layer and the meshed ground plane in copper below, as designed; the gaps in the trap electrodes here are 5~$\mu$m, and the ground mesh is formed of 600~nm wires with 350~nm gaps along $x$ and 10~$\mu$m gaps along $y$.  A chip cross section is diagrammed at right, with approximate relevant dimensions labeled (``pSi'' is polysilicon and metal interconnect layers are labelled m1 through m8).  Vias shown between metal layers are only representative.\label{figure:process}}

\end{figure}

%Designing into the process can also potentially lead to additional back-end processing utilizing the high-resolution foundry fabrication as a template for more %specialized, non-CMOS components, if desired.

Devices were fabricated on $3\times3$~mm$^{2}$ die (Fig.~\ref{figure:process}) on a shared, 300-mm, multi-project wafer produced in a 90-nm CMOS process operated by IBM (9LP process designation).  This process is primarily utilized for dense, high-performance digital circuits, and the trap die was one of many designs fabricated in parallel on the same wafer.  The process allows for patterning of 8~copper interconnect layers, along with the top aluminum pad layer (right panel of Fig.~\ref{figure:process}).  This $1.3$~$\mu$m thick pad layer was used for the trap electrodes, and a copper layer (m5) approximately 4~$\mu$m below the aluminum layer's bottom surface and 2~$\mu$m above the silicon substrate was used to form a ground plane under the extent of one of the traps. Due to metal density constraints arising from chemical-mechanical polishing steps applied to these layers, this ground plane was patterned as a mesh of 600~nm strips separated by 350~nm along the $x$ direction and 10~$\mu$m along the $y$ direction (see Fig.~\ref{figure:process}). Metal vias connect this copper ground plane to the center electrode of the trap through the upper metal layers m6--m8.

\begin{figure}[tbp]
\includegraphics[width=\columnwidth]{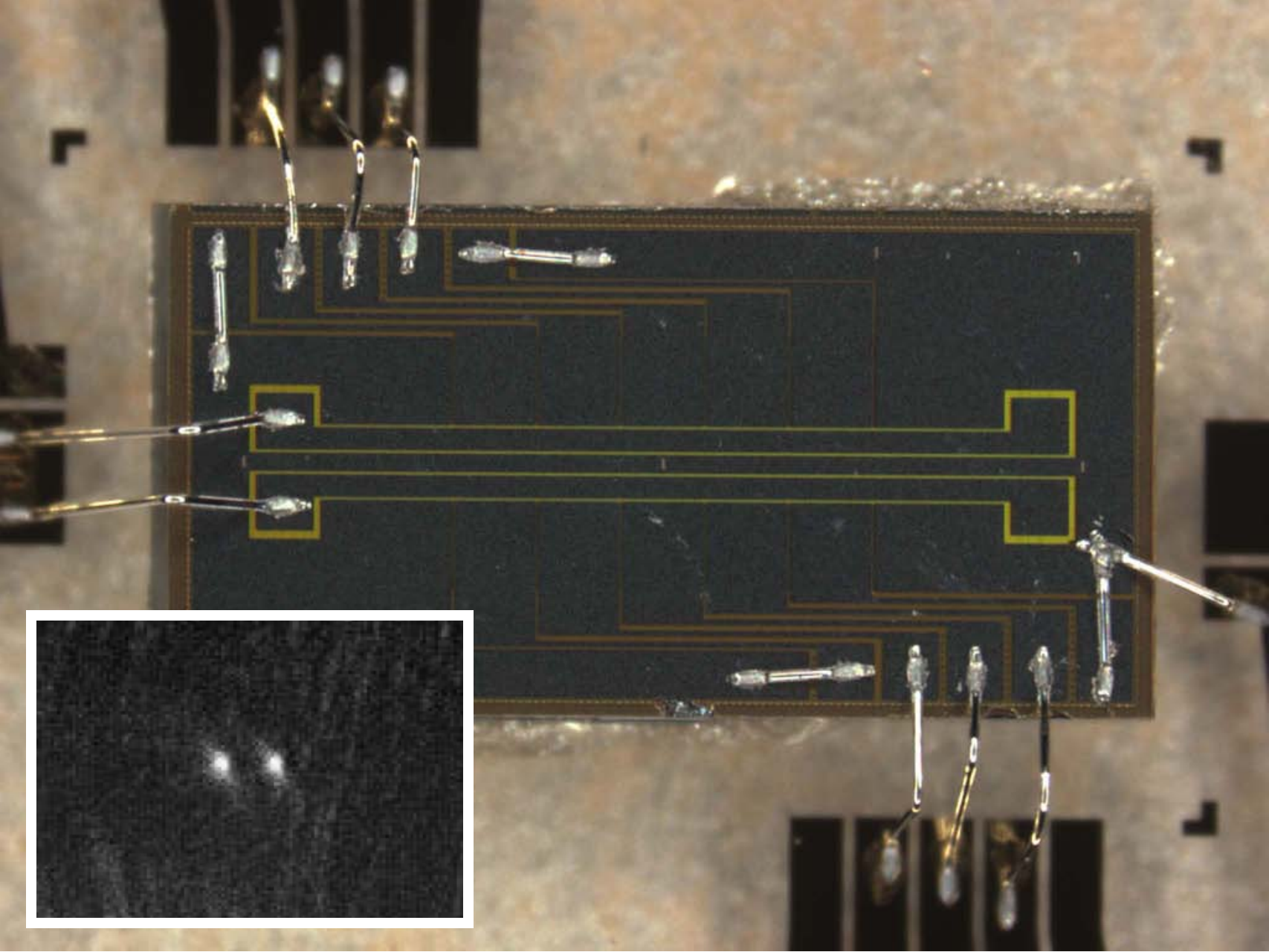}
\caption{(Color online) Micrograph of trap chip diced from die and mounted on the sapphire interposer of a cryogenic vacuum system.  Aluminum wirebonds are used to make contact from the aluminum trap electrodes to the gold interposer leads.  The chip is $2.5$~mm long and $1.2$~mm wide.  The inset shows two ions trapped 50~$\mu$m from the surface of the trap chip.  The ions are approximately 5~$\mu$m apart.\label{figure:mountedchip}}

\end{figure}

We designed and tested linear radio-frequency (RF) surface-electrode\cite{NIST:SET:QIC:05} Paul traps that confine ions 50~$\mu$m from the electrode surface.  The trap has a ``five-wire'' trap geometry, with two RF electrodes symmetric about the trap axis. Segmented dc control electrodes are routed to the corners of the trap chip to prevent wirebonds from obstructing laser access (see Fig.~\ref{figure:mountedchip}). This design offers flexibility to create various trapping potentials and allows scalability to multi-zone traps with complex geometries. Other advantages include comparative ease in selecting control voltages, and relatively narrow RF electrodes, allowing for lower capacitive coupling and RF power dissipation in the trap. Some inhomogeneity in RF field along the trap axis is anticipated due to the short (2~mm) electrode length, but effects on trapping are expected to be negligible.

%We predict a micromotion (ion motion at the RF frequency) amplitude due to edge effects of order 10~nm at an axial distance of 0.5~mm from the center of %the trap.

A possible limitation to the use of high-resolution CMOS processing is breakdown at large applied potentials, especially since typical ion trap voltage amplitudes are significantly higher than those used in CMOS electronics.  However, for up to 200~V static bias applied, the leakage current was below 10~pA, and no sudden increase corresponding to a dielectric breakdown was observed.  We performed these tests at room temperature (in a high-dielectric-strength fluid to prevent air breakdown), applying the potential between one of the RF electrodes and either the ground plane or one of the adjacent dc electrodes in both types of trap.

After commercial foundry fabrication, trap chips (diced from the full die) approximately $2.5\times1.2$~mm$^{2}$ were bonded to a larger interposer to interface with the cold stage and wiring of a cryogenic vacuum system that allows for variation of the trap-chip temperature\cite{PhysRevA.86.013417}.  Using a quarter-wave helical resonator, a $43$~MHz RF signal of approximately 100~V amplitude was applied to the trap electrodes to produce radial trap frequencies of approximately $4$--$5$~MHz, and an axial potential with frequencies near 1~MHz was produced by application of dc potentials of up to approximately 30~V to the segmented control electrodes.  We load $^{88}$Sr$^{+}$ ions by accelerating precooled Sr atoms from a magneto-optical trap toward the ion trap where they are photo-ionized and Doppler cooled\cite{PhysRevA.86.013417}.  Although not measured precisely in this work, loading efficiency into these traps is similar to more conventionally fabricated surface-electrode traps using the same loading method.

Traps without a ground plane displayed significant laser-induced photo-effects due to the excitation of carriers in the silicon by scattered light used for atom photoionization (PI, 405~nm) and ion Doppler cooling (422~nm).  During trap loading, this manifested itself as variation of the RF voltage amplitude on the trap electrodes due to varying impedance of the trap when the 405-nm PI light was on.  The effects on the trapping potential were visible as ion motion synchronized to the PI light switch state.  We observed no photo-effects in traps with a ground plane.  Traps without a ground plane also exhibit strong trap-temperature-dependent nonlinearities in the resonance response of the voltage-step-up resonator.  A ground plane reduces RF leakage into the silicon substrate sufficiently to eliminate this effect, such that we observed stable trapping for chip temperatures from 300~K down to 8~K.  We noticed slightly more power dissipation in the foundry traps than in traps fabricated from gold or niobium on sapphire for similar RF voltage amplitude\cite{PhysRevA.89.012318}, most likely due to higher dissipation in the metals or dielectrics.  Ion lifetimes of more than an hour were observed in the presence of Doppler cooling light, equivalent to the best lifetimes seen in other traps measured in this vacuum system.

Excess micromotion (ion motion at the RF drive frequency) is caused by static electric fields that displace the ion from the RF null and can lead to ion heating.  We compensate for this micromotion using the standard method of applying an additional opposing static field. Typical stray field values are on the order of 500~V/m here and appear stable over days.  Although silicon oxide dielectric is exposed at the locations of gaps in the electrodes and may charge due to laser-induced photo-electron production, the stray field's stability suggests another cause.  Wirebonds, which are asymmetric with respect to the ion location and also closer to the ion here than in the case of larger trap chips, may be responsible for the steady stray electric field.  The use of through-silicon-via technology can eliminate wirebonds from the chip surface, as has recently been demonstrated for surface-electrode ion traps\cite{gtri:invacuumelectronics:2014}.

When compared to single-metal-layer traps (SMLTs) on sapphire substrates, these traps exhibit increased scatter of laser light, possibly due to higher as-deposited roughness of the aluminum layer.  We examined the trap-electrode surface using atomic force microscopy and measured an RMS roughness of 35~nm, significantly larger than the 2~nm we have measured on SMLTs.  Scatter from the surface can be reduced by focusing laser beams to a smaller diameter at the trap.

\begin{figure}[tbp]
\includegraphics[width=\columnwidth]{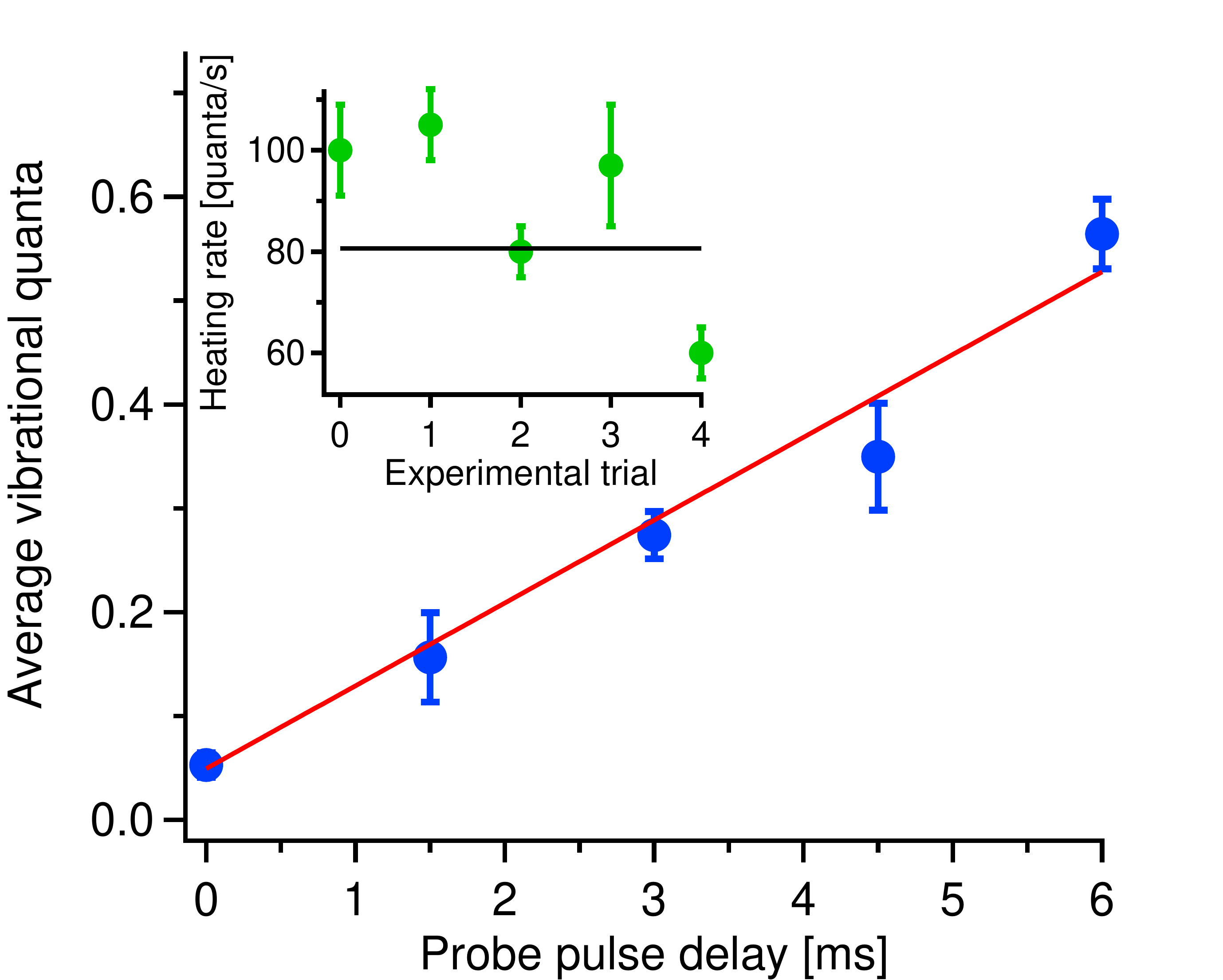}
\caption{(Color online) Representative measurement of heating rate in a CMOS-foundry-fabricated ion trap.  Average occupation of the axial mode of vibration in the linear trap is plotted as a function of delay time after preparation in the ground state at a trap temperature of 8.4~K.  A linear fit (line shown) gives a heating rate for these data of 80(5)~quanta/s where the uncertainty is due to statistical errors propagated through the fit. An average of five such measurements gives a heating rate of 81(9)~quanta/s where the uncertainty is due to run-to-run variability.  The inset shows all five measurements with a line indicating the weighted average value.\label{figure:heating}}

\end{figure}

Trapped-ion multi-qubit quantum operations can be limited by electric field noise that heats the ions' shared vibrational modes in the trap, reducing gate fidelity\cite{NIST:ExpIssueswithIons:JresNIST:98,nist:muwavetechniques:2013}.  Anomalously large heating rates caused by unknown noise sources have been seen in every trapped-ion experiment that has examined motional-state heating.  This is particularly noticeable in small microfabricated traps as the heating rate appears to scale as $1/d^{4}$ for an ion a distance $d$ from a trap electrode surface.  It is therefore important to characterize the heating rate in potentially scalable trap technologies.

Using the dipole-forbidden $S_{1/2}\to D_{5/2}$ transition in $^{88}$Sr$^{+}$, we performed resolved-sideband cooling to prepare the ion in the ground state (average occupation $\bar{n}\approx0.05$) of the 1.3~MHz axial vibrational mode and then measured the heating rate using sideband amplitude spectroscopy on this transition after a varying delay\cite{PhysRevA.89.012318}.  Results of one such measurement are presented in Fig.~\ref{figure:heating} for a chip temperature of 8.4~K.  Five measurements were recorded over a few days for nominally the same conditions; the average heating rate is 81(9)~quanta/s.   When scaled by $1/d^{4}$ to compare traps of different sizes, this heating rate is lower than that reported in any other trap fabricated on a silicon substrate\cite{britton:nist:2008,SMIT:QIC:2009,Oxford:CaCharging:APB:11,npl:monolithictrap:2012,vittorini:043112,Innsbruck:CryoSiTrap:arXiv:14}.  Motional heating at this level would lead to an error of less than $10^{-2}$ in a 100~$\mu$s two-ion-qubit gate, below the fault-tolerance threshold for large scale quantum computing with surface-code error-correction schemes\cite{fowler:surfacecodereview:PRA09}.

%Compared with all previously demonstrated ion traps (on any substrate), the traps described here allow for significantly more complexity and better %resolution.

We have shown basic functionality for quantum processing using a fabrication process, without modification, that has enabled scaling to billions of transistors.  This is the first demonstration, in any physical implementation, of quantum computing hardware co-fabricated with scalable classical computing hardware.  The fabrication of advanced CMOS and photonic technology on the trap chip, including the extensive existing libraries of integrated circuits for digital logic and memory, offers a straightforward path to scalable, local optical and electronic control and readout of trapped-ion arrays.  The demonstration of stable trapping and low electric-field noise in a foundry-process trap is therefore an initial step toward integration of the required classical computing and photonic devices for useful, large-scale quantum processing with trapped ions.  

%The fabrication of advanced CMOS and photonic technology on the trap chip, including the extensive, previously developed library of integrated circuits for %digital logic, microprocessors, and memory, is a straightforward path to fast, compact, and power-efficient local optical and electronic control and readout of %trapped-ion arrays. 

\begin{acknowledgments}
We thank Peter Murphy, Chris Thoummaraj, and Karen Magoon for assistance with ion-trap-chip packaging at Lincoln Laboratory.  K.K.M. and R.J.R acknowledge funding from DARPA MTO, NSF iQuISE, and a DOE Science Graduate Fellowship.  I.L.C. acknowledges funding from IARPA.  The work at Lincoln Laboratory is sponsored by the Assistant Secretary of Defense for Research \& Engineering under Air Force contract number FA8721-05-C-0002. Opinions, interpretations, conclusions, and recommendations are those of the authors and are not necessarily endorsed by the United States Government.
\end{acknowledgments}

\bibliography{jc_bib1,jc_bib2,IonLoading_PRAfinal}

\end{document}